\documentclass[prd,onecolumn]{revtex4}
\pdfoutput=1
\usepackage{amssymb,latexsym}
\usepackage{amsmath,amsbsy,bbm}
\usepackage{ifpdf}
\usepackage{epsfig,bm}
\usepackage{graphicx,comment}
\usepackage{color}
\usepackage{soul}
\usepackage{mathtools}
\usepackage{comment}
\usepackage[normalem]{ulem}
\begin{document}

\title{A Bond weighted tensor renormalization group study of the $q$-state ferromagnetic Potts models on the square lattice}

\author{Yuan-Heng Tseng}
\affiliation{Department of Physics, National Taiwan Normal University,
88, Sec.4, Ting-Chou Rd., Taipei 116, Taiwan}
\author{Shang-Wei Li}
\affiliation{Department of Physics, National Taiwan Normal University,
	88, Sec.4, Ting-Chou Rd., Taipei 116, Taiwan}
\author{Fu-Jiun Jiang*}
\affiliation{Department of Physics, National Taiwan Normal University,
88, Sec.4, Ting-Chou Rd., Taipei 116, Taiwan\\
\\
*Corresponding author. Email: fjjiang@ntnu.edu.tw\\
Contributing authors' emails: yhtsengr@gmail.com (Yuan-Heng Tseng) and wbby0623@gmail.com (Shang-Wei Li)
}

\begin{abstract}
  It is known rigorously that the phase transition of the $q$-state ferromagnetic Potts model
  on the square lattice is second order for $q=4$. Despite this fact, some observables of
  the $q=4$ model show features of a first-order phase transition. For example, negative
  peak appears for the quantity of 
  Binder ratio $Q_2$ of this model. Such a non-monotonic behavior of $Q_2$ is typically a consequence of
  phase coexistence, hence is served as a signal of a first-order phase transition. In particular,
  the negative peak should diverge with linear system size $L$ squared. Since the mentioned divergence phenomenon is
  not observed for the 4-state Potts model, the scenario of a first-order phase transition for this model is ruled out. Interestingly, a recent large scale Monte Carlo investigation of the 4-state Potts model observes that the two-peak structure of the energy density distribution becomes more noticeable when $L$ increases. This finding indicates the signal of coexistence of phases is getting stronger with $L$. Due to these unusual critical
  behaviors, here we study the energy density $E$ and the specific heat $C_v$ of the 4-state Potts model on the square lattice using the technique
  of bond weighted tensor renormalization group (BWTRG). For a comparison purpose, $q=2$ and $q=5$ ferromagnetic Potts models on the square lattice are investigated using the same method as well. Remarkably, our results do imply there may be a
  small energy gap for $q=4$ model. While the appearance of the mentioned small energy gap can be explained plausibly and it will disappear with a more sophisticated investigation,
  our finding suggests that whether a message of a first-order phase transition is genuine or
  is an artificial effect requires further and detailed investigations.
  
\end{abstract}

\maketitle

\section{Introduction}

The $q$-state ferromagnetic Potts models have been studied extensively both with
analytic and numerical methods \cite{Pot52,Bax73,Bax78,Hin78,Blo79,Wu82,Bax82,Car86,Fuk89,Hu89,Lee91,Alv91,Jan97,Sal97,Dum17}.
At the moment, it is demonstrated rigorously that in spatial dimension two, the phase transitions of $q < 5$ ferromagnetic
Potts model on the square lattice are second order \cite{Bax73,Dum17}.

4-state Potts model on the square lattice is rather special and interesting.
Apart from receiving multiplicative logarithmic corrections, certain features of
first-order phase transition appear in its critical behaviors.

For a first-order phase transition in spatial dimension two,
due to the coexistence of phases,
the quantity Binder ratio $Q_2$ develops negative peak approaching the critical point with increasing linear system sizes $L$. In addition, the peak diverges with $L^2$.
In reality, the $Q_2$ of the 4-state Potts model on the square lattice does
develop negative peak. However, the peak does not grow with $L^2$. Hence, the first-order phase transition scenario is ruled out for the 4-state Potts model, and in
the literature a second-order phase transition with such an exotic behavior is
often referred to as the pseudo-first-order phase transition \cite{Jin12,Jin13}.

Interestingly, a recent large scale Monte Carlo (MC) calculation in \cite{Pen24}
finds that for the energy density $E$ distribution of the 4-state ferromagnetic
Potts model on the square lattice, two-peak structure, which is a signal of a first order phase transition, appears and this phenomenon becomes more    
and more noticeable with increasing linear system sizes $L$. The largest $L$ considered in Ref.~\cite{Pen24} is $L=4096$,
and ideally one should perform calculations with larger lattices to examine
whether the sign of two-peak structure begins to diminish or stops getting stronger.  

Performing MC simulations with $L \gg 4096$ is very time consuming. As a result, to shed some light on the unexpected
observation in Ref.~\cite{Pen24}, in this study we carry out a tensor network (TN) calculation to explore the nature of
the phase transition of the 4-state ferromagnetic Potts model on the square lattice. Specifically, we apply the
bond weighted tensor renormalization group method (BWTRG) \cite{BWTRG} to compute the specific heat $C_v$ and energy density $E$ close to the critical temperature $T_c$
of the considered model. For a comparison purpose, the phase transitions of
$q=2$ and $q=5$ ferromagnetic Potts models on the square lattice are investigated by employing the same technique of BWTRG.

For $q=2$ and $q=5$ models, based on the outcomes obtained by BWTRG, one can conclude without ambiguity that the phase transitions of $q=2$ and $q=5$ models are second order and first order, respectively. 

Surprisingly, our BWTRG calculations (on a $L=2^{17}$ square lattice) lead to the finding that for the phase transition of $q=4$ model, there is a tiny,
but non-vanishing energy gap. In other words, even with the computation on lattices much larger than $L=4096$ (which is the one
used in Ref.~\cite{Pen24}), we arrive at the result that a characteristic of first-order phase transition still shows up
for the considered phase transition. Although the presence of the mentioned energy gap can be understood plausibly, in other words, this tiny energy gap disappears when a more sophisticated investigation is conducted,
our finding suggests that whether a message of a first-order phase transition is genuine or is an artificial effect
requires further and detailed investigations.

The rest of the paper is organized as follows. After the introduction,
the $q$-state ferromagnetic Potts model on the square lattice is described in Sec.~II briefly. 
We then demonstrate the obtained results in Sec.~III. In particular, the illusion of a non-vanishing energy gap
is presented for $q=4$ model.
Finally, Sec.~VI contains the discussions and conclusions of the our study.

\section{The considered model and observables}

The Hamiltonian $H_{\text{Potts}}$ of the $q$-state ferromagnetic Potts model on the square lattice studied here is expressed as \cite{Wu82}
\begin{equation}
\beta H_{\text{Potts}} = -\beta \sum_{\left< ij\right>} \delta_{\sigma_i,\sigma_j}.
\label{eqn}
\end{equation}
Here $\beta$ is the inverse temperature (Temperature is denoted by $T$ here), $\delta$ refers to the Kronecker function, and the Potts variable
$\sigma_i$ at each site $i$ takes an integer value from $\{1,2,...,q-1,q\}$.

In this study, we conduct a detailed investigation for the energy density $E$  and the specific heat $C_v$ as functions of the $T$ near the critical temperatures for $q=2$, 4 and 5. The specific heat takes the form
\begin{equation}
C_v = \frac{L^2}{T^2}\left(\langle E^2\rangle-\langle E \rangle^2 \right),
\end{equation}
here $L$ is the linear system size. 
$C_v$ can be obtained by differentiating $E$ as well. In this study, we use this later approach to calculate $C_v$.

\section{Method}

\begin{figure}
    \includegraphics[width=0.7\textwidth]{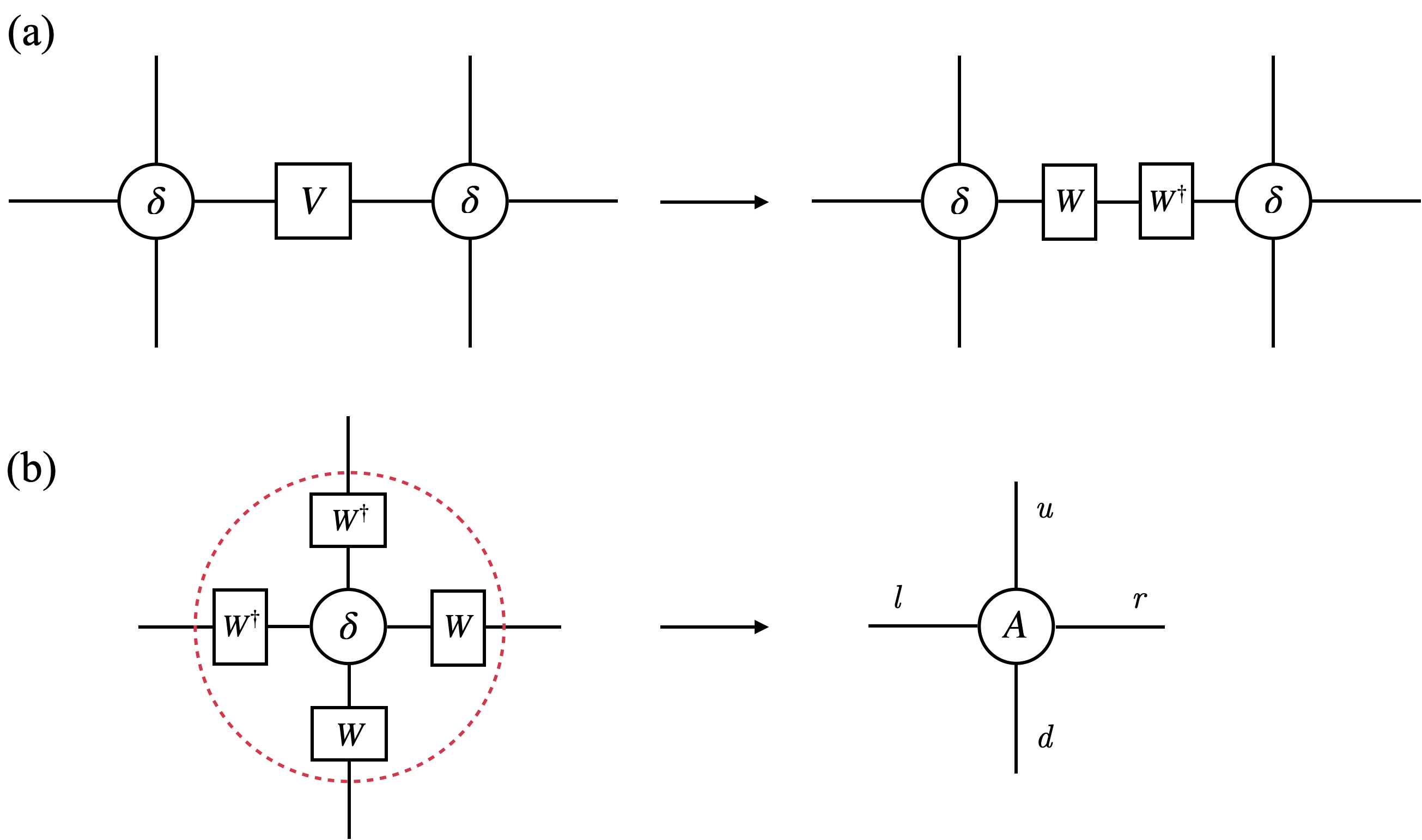}
	\caption{The construction of the local tensor $A$. (a) The local bond matrix $V$ is decomposed to $WW^{\dagger}$. (b) The local tensor $A$ is obtained by absorbing the matrices $W$ and $W^{\dagger}$ into a single site.}
	\label{fig_local_tensor}
\end{figure}

It is well known that the partition function of all classical statistical models with short-range interactions, including the Ising model and the Potts model, can be represented as tensor network models \cite{TNS}. For the $q$-state Potts model on a square lattice with periodic boundary conditions, the partition function can be expressed as
\begin{equation}
    Z = \text{Tr} e^{-\beta H_{\text{Potts}}}=\text{Tr}\Pi_{i}A_{l_i,u_i,r_i,d_i},
\end{equation}
where $A$ is a rank-4 local tensor with bond dimension $q$, the trace is to sum over all bond indices, and $(l_i, u_i, r_i, d_i)$ denote the bond indices connecting site $i$ from (left, up, right, down) directions, respectively. The local tensors $A$ can be obtained by factorizing the local bond matrices that connect two neighboring sites using the method of singular value decomposition SVD (or the eigendecomposition) \cite{Lev07}, see fig.~\ref{fig_local_tensor}. Each local bond matrix of the $q$-state Potts model is given by $V_{ij}=\left< \sigma_i \left|e^{\beta\delta_{\sigma_i,\sigma_j}}\right|\sigma_j \right>$, which is a $q$ by $q$ positive-definite matrix, and can be decomposed to $WW^{\dagger}$ through eigendecomposition. Here $W$ is a matrix and $W^{\dagger}$ is its Hermitian adjoint. The local tensors are then constructed by
\begin{equation}
    A_{l_i,u_i,r_i,d_i}=\sum_{\alpha,\beta,\gamma,\rho} \delta_{\alpha\beta\gamma\rho} W^{\dagger}_{l_i\alpha}W^{\dagger}_{u_i\beta}W_{r_i\gamma}W_{d_i\rho},
\end{equation}
where $\delta_{\alpha\beta\gamma\rho}$ is the rank-4 Kronecker delta function.

\begin{figure}
    \includegraphics[width=0.95\textwidth]{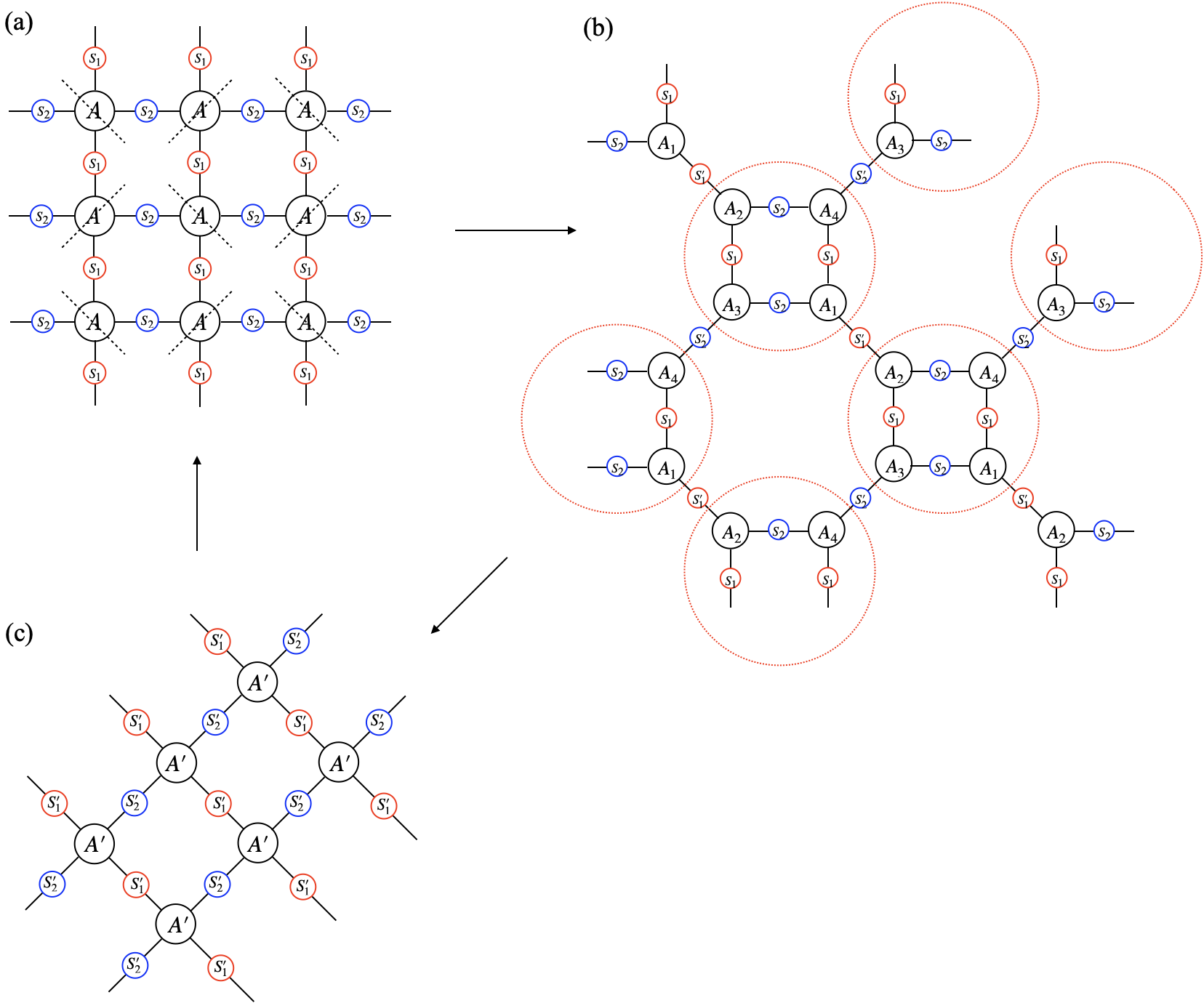}
	\caption{A single renormalization step of BWTRG. (a) Each local tensor $A$ is decomposed using SVD along a direction that depends on its position. (b) The new local tensors are obtained by contracting the corresponding tensors in the plaquette, while the new bond weights $S_1^{\prime}$ ($S_2^{\prime}$) between the plaquettes are given by $\sigma_1^k$ ($\sigma_2^k$). (c) The renormalized tensor network can be restored to the original one by rescaling with a factor of $1/\sqrt{2}$ and rotating by $\pi/4$.}
	\label{fig_BWTRG}
\end{figure}

After representing the partition function as a tensor network model composing of local tensors $A$, one can calculate the partition function by contracting the entire network. This cannot be done exactly because the bond dimension will increase after each contraction. However, tensor renormalization group approaches \cite{Lev07,iTEBD,CTMRG,Xie12,ATRG,BWTRG} address this problem by truncating the bond dimension to a manageable number during the renormalization. In this paper, we empoly the bond-weighted tensor renormalization group (BWTRG) method \cite{BWTRG} for tensor renormalization. In BWTRG, there are additional bond weights (rank-2 tensors) on the edges of the tensor network when compared to the standard tensor renormalization group method TRG \cite{Lev07}. Therefore, in addition to updating the local tensors, one needs to update the bond weights at each renormalization step. First, we decompose each rank-4 local tensor into two rank-3 tensors and one rank-2 tensor using SVD along one of two possible directions
\begin{equation}
    A_{l_i,u_i,r_i,d_i} \approx \sum_\alpha^D U_{1(l_i,u_i),\alpha} \sigma_{1\alpha,\alpha} V_{1\alpha,(r_i,d_i)}^{\dagger},
\end{equation}
\begin{equation}
    A_{l_i,u_i,r_i,d_i} \approx \sum_\alpha^D U_{2(u_i,r_i),\alpha} \sigma_{2\alpha,\alpha} V_{2\alpha,(l_i,d_i)}^{\dagger},
\end{equation}
where $D$ is the cutoff bond dimension. Then, the initial bond weights $S_1$ and $S_2$ (identity matrices), which are connected to the nearest plaquettes, are replaced with $\sigma_1^k$ and $\sigma_2^k$, respectively. In this case, the current algorithm differs from TRG due to the introduction of the hyperparameter $k$. One notices that by setting $k=0$, the original TRG is recovered. The corresponding tensors at the vertices of the plaquette now become
\begin{equation}
    A_{1(l_i,u_i),\alpha}=U_{1(l_i,u_i),\alpha}\sigma_{1\alpha,\alpha}^{(1-k)/2},
\end{equation}
\begin{equation}
    A_{2\alpha,(r_i,d_i)}=\sigma_{1\alpha,\alpha}^{(1-k)/2}V_{1\alpha,(r_i,d_i)}^{\dagger},
\end{equation}
\begin{equation}
    A_{3(u_i,r_i),\alpha}=U_{2(u_i,r_i),\alpha}\sigma_{2\alpha,\alpha}^{(1-k)/2},
\end{equation}
\begin{equation}
    A_{4\alpha,(l_i,d_i)}=\sigma_{2\alpha,\alpha}^{(1-k)/2}V_{2\alpha,(l_i,d_i)}^{\dagger}.
\end{equation}
Finally, the updated local tensors and bond weights are constructed as follows:
\begin{equation}
    A_{l_i,u_i,r_i,d_i}^{\prime}=\sum_{\alpha\beta\gamma\rho}A_{3(\alpha,\beta),l_i}S_{1\alpha,\alpha}A_{2u_i,(\gamma,\alpha)}S_{2\gamma,\gamma}A_{4r_i,(\gamma,\rho)}S_{1\rho,\rho}A_{1(\beta\rho),d_i}S_{2\beta,\beta},
\end{equation}
\begin{equation}
    S_1^{\prime}=\sigma_1^k,
\end{equation}
\begin{equation}
    S_2^{\prime}=\sigma_2^k.
\end{equation}
The renormalized tensor network can be restored to the original one by rotating $\pi/4$ and rescaling with a factor of $1/\sqrt{2}$. A single renormalization step of BWTRG is illustrated in fig.~\ref{fig_BWTRG}. In \cite{BWTRG}, the authors reported that the optimal value of $k$ is $-0.5$. The error in the free energy is minimized at this value, and the stationary condition of the fixed point tensor also supports it. The bond weights in BWTRG act as a mean-field environment \cite{TNS}, and are used to avoid the effects of the loop entanglement 
\cite{CDL,Gilt} during tensor renormalization. In this paper, we will fix the value of $k$ to this optimal value. 

The initial tensor network contains $2^{N/2}\times2^{N/2}$ local tensors. After $N$ steps of renormalization, only one renormalized local tensor remains. The partition function is given by tracing out all indices. Once the partition function is obtained, various thermodynamic quantities can be derived from it, including the energy density $E$ and the specific heat $C_v$.
We will use the automatic differentiation technique \cite{autodiff} to calculate $E$ and $C_v$. 

\section{Numerical Results}

\subsubsection{Results of 2-state ferromagnetic Potts model}

As a test case, we have applied the technique of BWTRG to calculate $C_v$ and $E$ near the critical temperature
$T_c$ of the 2-state ferromagnetic Potts model on the square lattice.

\begin{figure}
	\hbox{
	\includegraphics[width=0.5\textwidth]{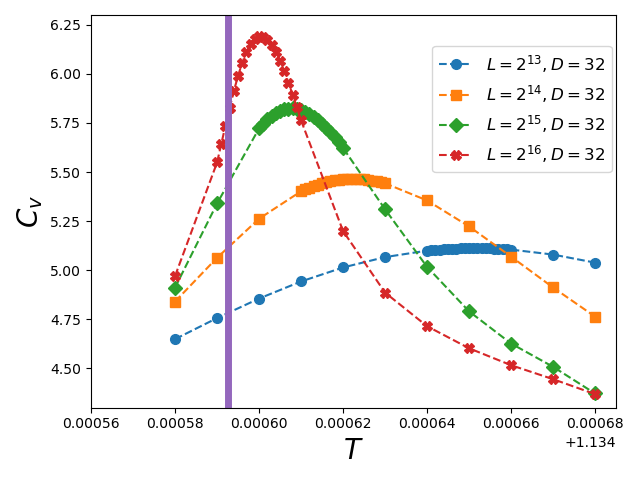}~~~~
	\includegraphics[width=0.5\textwidth]{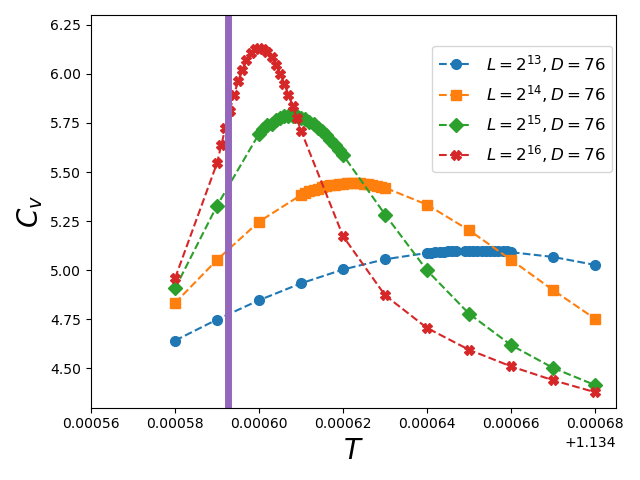}
}
	\caption{The $T$-dependence of $C_v$ for the 2-state Potts model on the square lattice. The bond dimension $D$
		for the left and the right panels are 32 and 76, respectively.}
	\label{fig1}
\end{figure}

Fig.~\ref{fig1} shows the specific heat $C_v$ as functions of $T$ for several 
steps of performing renormalization group contraction. The linear system size $L$
is determined by how many RG steps are carried out. For example, after the $N$ RG steps, $L$ is $2^{N/2}$. The bond dimension $D$
for the left and the right panels of fig.~\ref{fig1} are 32 and 76, respectively. Finally, the vertical solid lines in both panels are the theoretical $T_c$. As can be seen from
the figure, the $T$ corresponding to the peaks approach $T_c$ as
$L$ increases.

If the phase transition is first order, then the peaks of $C_v$ should scale
with $L^2$. This is not what's been observed in fig.~\ref{fig1}. As a result,
the phase transition of the 2-state Potts ferromagnetic model on the square lattice is not first order.

\begin{figure}
	\hbox{
		\includegraphics[width=0.5\textwidth]{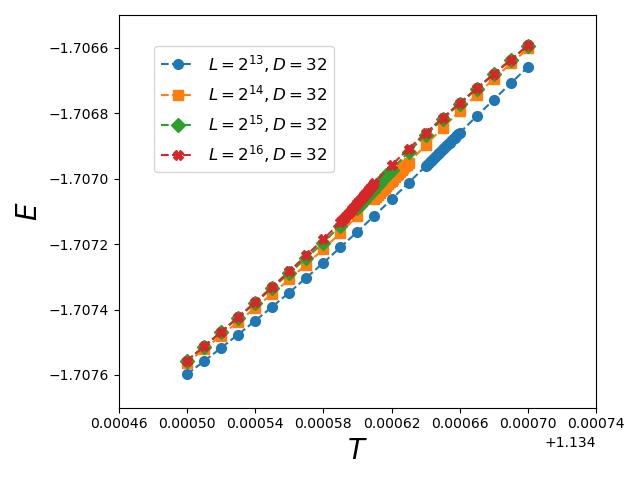}~~~~
		\includegraphics[width=0.5\textwidth]{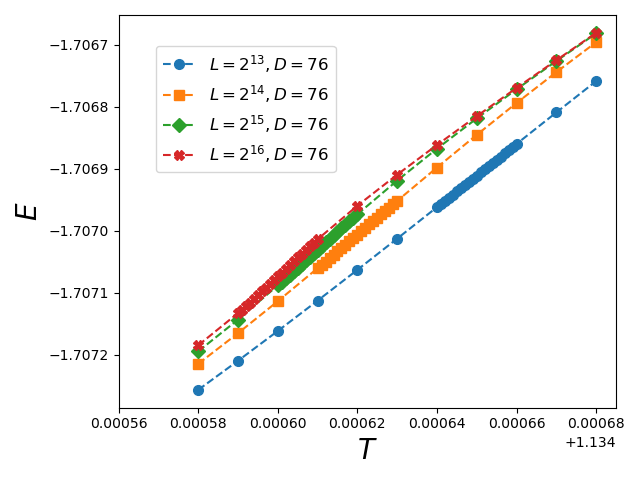}
	}
	\caption{The $T$-dependence of $E$ for the 2-state Potts model on the square lattice. The bond dimension $D$
		for the left and the right panels are 32 and 76, respectively.}
	\label{fig2}
\end{figure}

Fig.~\ref{fig2} demonstrates the energy density $E$ as functions of $T$ obtained by performing several RG steps. The bond dimension $D$
for the left and the right panels of fig.~\ref{fig1} are 32 and 76, respectively. The results appearing in fig.~\ref{fig2} provide convincing evidence that there is no gap in the energy density $E$. 

Based on the outcomes depicted in figs.~\ref{fig1} and \ref{fig2},
one concludes that the phase transition associated with the 2-state ferromagnetic Potts model is second order as one expects.

We would like to point out that for a second-order phase transition in spatial dimension two, $C_v(L) \sim L^{\alpha/\nu}$ close to $T_c$. The peaks of $C_v$ for the system sizes shown in fig.~\ref{fig2} increase only slightly as one goes from
$L=2^{15}$ to $2^{16}$.
This leads to the claim that $\alpha = 0$. This is consistent with the theoretical prediction ($\alpha = 0$ for 2-state ferromagnetic
Potts model on the square lattice).

\subsubsection{Results of the 4-state ferromagnetic Potts model}

After confirming the phase transition of the 2-state ferromagnetic Potts model is second order using the technique of BWTRG, we turn to investigate
the 4-state Potts model on the square lattice. Similar to the 2-state Potts model, 
we have performed the calculations with two values of bond dimension, namely
$D=32$ and $D=76$.

\begin{figure}
		\includegraphics[width=0.5\textwidth]{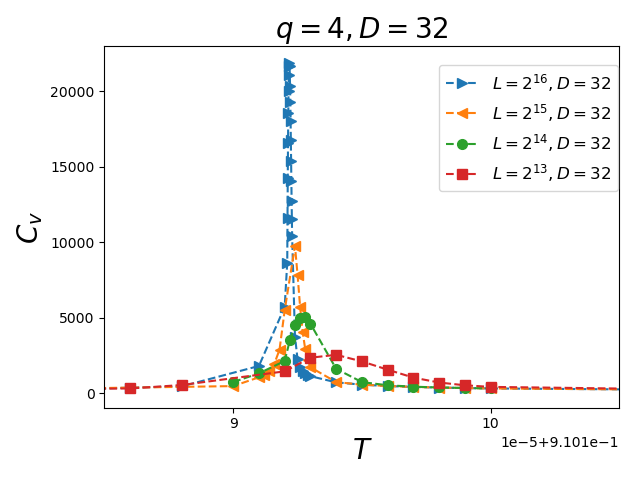}
	\caption{The $T$-dependence of $C_v$ for the 4-state Potts model on the square lattice. The bond dimension $D$ is 32.}
	\label{fig3}
\end{figure}

$C_v$ as functions of $T$ for $L = 2^{13}, 2^{14}, 2^{15}, 2^{16}$ are shown in fig.~\ref{fig3}. As can be seen from the figure, the maximum values of $C_v$ 
for these considered $L$ do not scale with $L^2$. Therefore it is unlikely that
the associated phase transition is first order.

It is known that for a second-order phase transition, the maximum values of $C_v$ at various $L$ scale with $L^{\alpha/\nu}$, and 
theoretically, $\alpha/\nu = 1$ for the 2D ferromagnetic four-state Potts model on the square lattice.

\begin{figure}
		\includegraphics[width=0.5\textwidth]{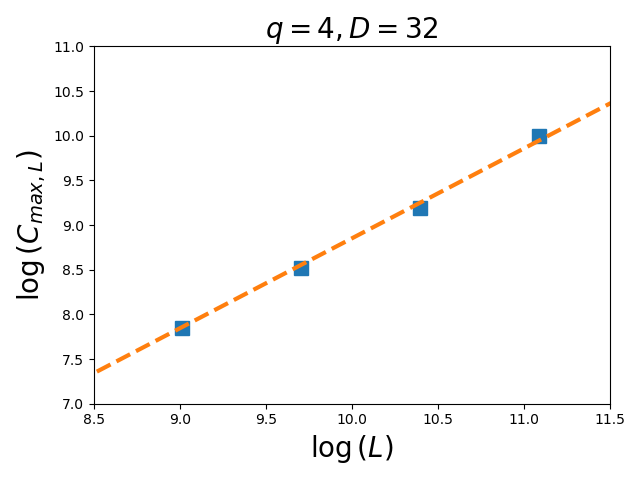}
	\caption{The $\left(\ln L\right)$-dependence of $\ln \left(C_{\text{max},L}\right)$ for the 4-state Potts model on the square lattice. The bond dimension $D$ is 32.}
	\label{fig3.1}
\end{figure}

We have extracted the peak values $C_{\text{max},L}$ of fig.~\ref{fig3} for various $L$, and 
$\ln \left(C_{\text{max},L}\right)$ as a function of $\ln L$ is shown in fig.~\ref{fig3.1}. The straight line in fig.~\ref{fig3.1} has a slope of 1.005. This implies that the associated $\alpha/\nu$ for the considered phase transition is around 1.
This is consistent with the theoretical prediction. The outcome demonstrated in fig.~\ref{fig3.1} provides a convincing evidence that the phase transition of
the 2D four-state ferromagnetic Potts model on the square lattice is second order and the
related (combination of) critical exponents agrees with its analytic prediction. 

\begin{figure}
	\hbox{
	\includegraphics[width=0.5\textwidth]{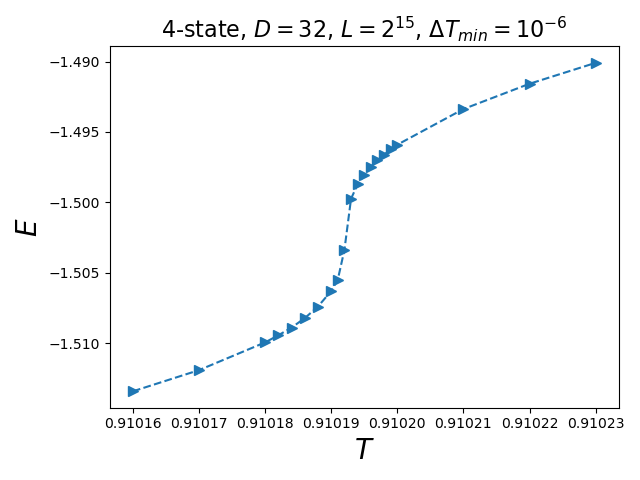}
		\includegraphics[width=0.5\textwidth]{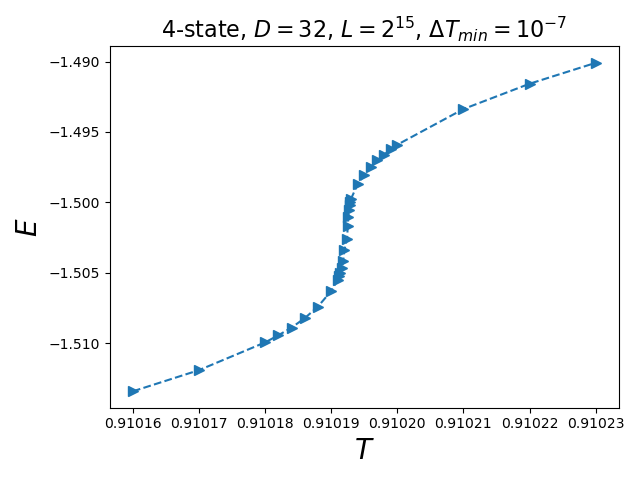}
}
	\caption{The $T$-dependence of $E$ for the 4-state Potts model on the square lattice. $L = 2^{15}$ and the bond dimension $D$ is 32.}
	\label{fig4}
\end{figure}

$E$ as functions of $T$ close to $T_c$ for $L = 2^{15}$ are shown fig.~\ref{fig4}.
The separation between two consecutive temperatures $\Delta T_{\text{min}}$ for the left and the right panels are given by $10^{-6}$ and $10^{-7}$, respectively.

Interestingly, the left panel of fig.~\ref{fig4} indicates there is a small energy gap. However, this small energy gap disappears when one considers $\Delta T_{\text{min}} = 10^{-7}$ in the calculations. 
 
\begin{figure}
	\hbox{
		\includegraphics[width=0.5\textwidth]{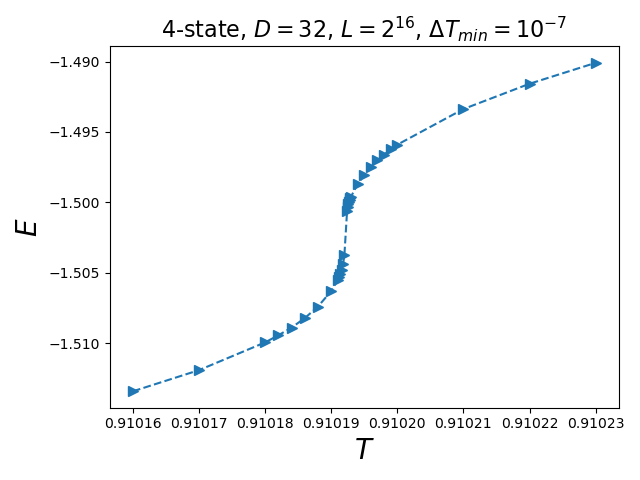}
		\includegraphics[width=0.5\textwidth]{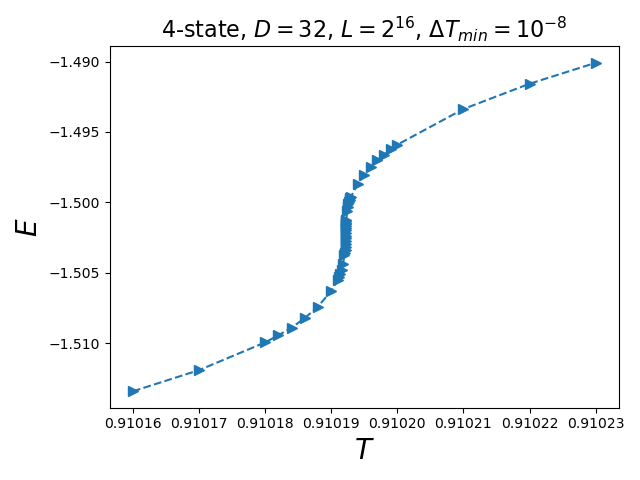}
	}
	\caption{The $T$-dependence of $E$ for the 4-state ferromagnetic Potts model on the square lattice. The bond dimension $D$ is 32.}
	\label{fig5}
\end{figure}

$E$ as functions of $T$ close to $T_c$ for $L = 2^{16}$ with $\Delta T_{\text{min}} = 10^{-7}$ and $\Delta T_{\text{min}} = 10^{-8}$ are shown in the left and the right panels of fig.~\ref{fig5}.

Similar to the results appearing in fig.~\ref{fig4}, a tiny energy gap shows up
in the left panel which has $T_{\text{min}} = 10^{-7}$. This energy gap is gone
when $\Delta T_{\text{min}}$ used is $10^{-8}$, i.e. the right panel of fig.~\ref{fig4}.

Based on the outcomes shown in figs.~\ref{fig4} and \ref{fig5}, it is likely
that the energy gaps occurring in the left panels of these figures are due to an artifact associated with $\Delta T_{\text{min}}$. Specifically, energy gap may show up if $\Delta T_{\text{min}}$ is not small enough. By considering smaller $\Delta T_{\text{min}}$, the energy gaps are gone, hence the true nature of the second phase transition for the 4-state antiferromagnetic Potts model emerges then.

\begin{figure}
	\hbox{		\includegraphics[width=0.33\textwidth]{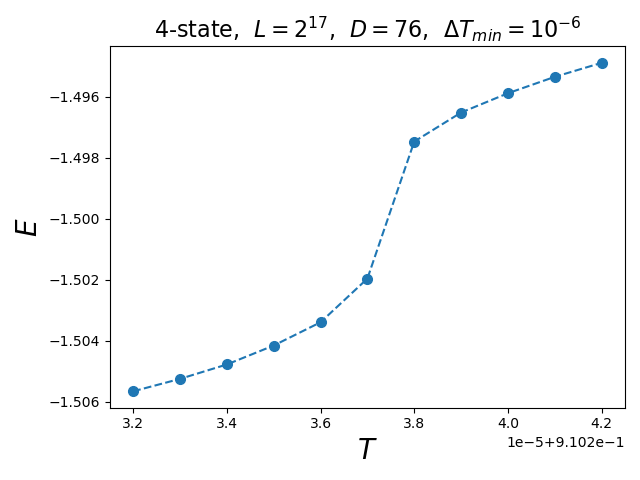}		\includegraphics[width=0.33\textwidth]{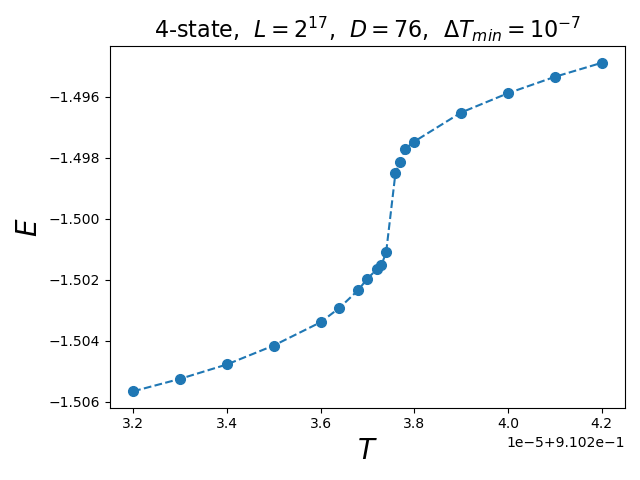}        \includegraphics[width=0.33\textwidth]{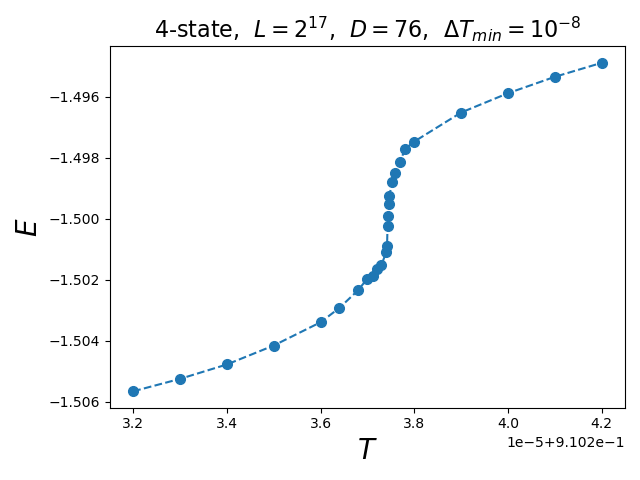}}
	\caption{The $T$-dependence of $E$ for the 4-state ferromagnetic Potts model on the square lattice. The bond dimension $D$ is 76.}
	\label{fig51}
\end{figure}

To confirm the statement outlined in the previous paragraph, we have conducted calculations with $D=76$, $L=2^{17}$ and the associated outcomes are depicted in fig.~\ref{fig51}. The left, the middle, and the right panels are related to $\Delta T_{\text{min}} = 10^{-6}, 10^{-7}$, and $10^{-8}$, respectively. Clearly, with larger $D$ and $L$, the conclusion drawn previously is still valid, namely, a energy gap shows up for large $\Delta T_{\text{min}}$ and it disappears when one considers smaller $\Delta T_{\text{min}}$.

\subsubsection{Results of the 5-state ferromagnetic Potts model}

To verify that the disappearance of the tiny energy gaps shown in fig.~\ref{fig4} and \ref{fig5} is generic, we have performed a similar calculation for the 5-state ferromagnetic Potts model on the square lattice. Rigorously, this model has a weak first-order phase transition which has a long correlation length about 2500 \cite{Bud93}. 

\begin{figure}
	\hbox{
		\includegraphics[width=0.5\textwidth]{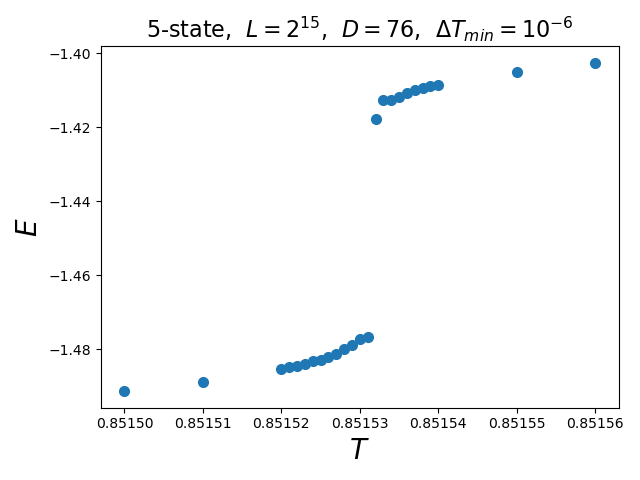}
		\includegraphics[width=0.5\textwidth]{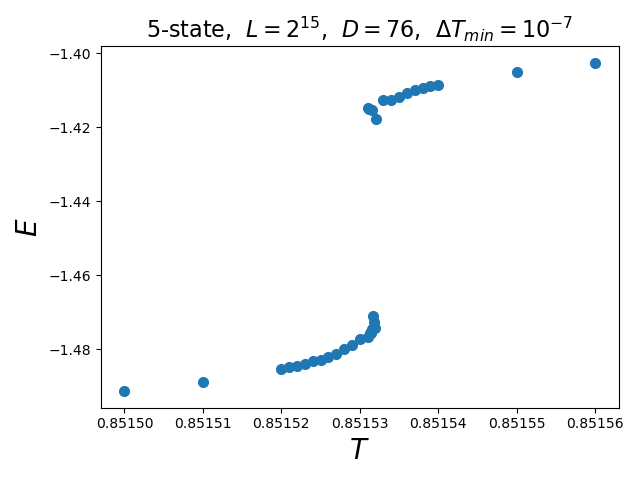}
	}
	\caption{The $T$-dependence of $E$ for the 5-state Potts model on the square lattice. The bond dimension $D$ is 76.}
	\label{fig6}
\end{figure}

Clearly, the energy gaps shown in both panels of fig.~\ref{fig6} are much larger than those appearing in the left panels of figs. \ref{fig4} and \ref{fig5}.
In particular, unlike the scenarios of figs.~\ref{fig4} and \ref{fig5},
the outcomes of fig.~\ref{fig6} indicate that the energy gap does not diminish when the magnitude of $\Delta T_{\text{min}}$ decreases. This shows the fundamental
difference between the phase transitions of 4- and 5-state ferromagnetic Potts models on the square lattice.

\section{Discussions and Conclusions}

In this study, we investigate the nature of the phase transitions of the 2-, 4-, and 5-state ferromagnetic Potts models on the square lattice with the
method of BWTRG. In particular, the temperature $T$ dependence of the energy density $E$ is explored. For the 2-state model, when the quantity $E$ is considered as a function of $T$, no discontinuous behavior is found.
This provides strong evidence to prove that the corresponding phase transition is second order.

It has been shown in the literature that analytically (and rigorously) the phase transition of 4-state ferromagnetic Potts model on the
square lattice is second order \cite{Bax73,Dum17}. Therefore, the characteristics of a first-order phase transition 
observed in Ref.~\cite{Pen24} (i.e. the two-peak structure of the energy density distribution becomes more obvious when one increases $L$) and
here for the 4-state ferromagnetic model is very puzzling. While the Monte Carlo method is exact,
it suffers systematic finite-size effects. The largest linear system size $L$ used in Ref.~\cite{Pen24} is 4096 which may not be sufficiently
large to capture the true nature of the considered phase transition. Indeed, $L=4096$ is a rather small system size in the BWTRG calculations.
Due to this, we extend the calculations done in Ref.~\cite{Pen24} by employing the technique of BWTRG. While with BWTRG we are indeed
able to obtain outcomes on lattices much larger than 4096, an energy gap is still observed which is again a message of a first-order phase transition.
We argue it is likely that this is due to the fact
that two consecutive values of temperature used in the calculations is not sufficiently small, and this statement is confirmed in figs.~\ref{fig4}, \ref{fig5} and \ref{fig6}.

Typically, one would conclude that a phase transition is first order based on the outcomes shown in the left panels of fig.~\ref{fig4} since the associated
$\Delta T_{\text{min}} = 10^{-6}$ is already very fine. It turns out that the gap appearing in fig.~\ref{fig4} would be gone once one uses $\Delta T_{\text{min}} = 10^{-7}$ to carry
out the BWTRG calculations. Our study suggests that in some cases, to determine a phase transition is first order or second order may be subtle.
In particular, deciding whether a feature of a first order phase transition is real or
is a false masquerade requires extremely careful investigations.

Finally, we would like to point out that the bond dimension $D$ should be as large as possible to avoid finite-size effects in a TRG calculation \cite{Ued23}. Here
we have performed the computations with $D=32$ and $D=76$. Both the resulting outcomes lead to the same conclusions. This
confirms that the data shown here and the claims drawn from them should be reliable.

\section*{Funding}\vskip-0.3cm
Partial support from National Science and Technology Council (NSTC) of
Taiwan is acknowledged (Grant numbers: NSTC 112-2112-M-003-016- and NSTC 113-2112-M-003-014-). 

\section*{Author Contributions}
F.-J. Jiang proposed and supervised the progress of the project, and wrote up the manuscript.
Y.-H. Tseng and S.-W. Li conducted the calculations and analyzed the data.

\section*{Conflict of Interest}
The authors declare no conflict of interest.

\section*{Data Availability Statement}
Data are available from the corresponding author
on reasonable request.

\bibliographystyle{unsrt}
\bibliography{reference}

\begin{thebibliography}{10}

\bibitem{Pot52}
R. B. Potts, Math. Proc. Camb. Philos. Soc. {\bf 48}, 106 (1952).

\bibitem{Bax73}
R. J. Baxter, J. Phys. C: Solid State Phys. {\bf 6}, L445 (1973).

\bibitem{Bax78}
R. J. Baxter, H. N. V. Temperley, and S. E. Ashley, Proc. R. Soc. Lond. A {\bf
  358}, 535 (1978).

\bibitem{Hin78}
A. Hintermann, H. Kunz, and F. Y. Wu, J. Stat. Phys. {\bf 19}, 623 (1978).

\bibitem{Blo79}
H.~W.~J. Bl$\ddot{o}$te and R.~H. Swendsen, Phys. Rev. Lett. {\bf 43}, 799
  (1979).

\bibitem{Wu82}
F. Y. Wu, Rev. Mod. Phys. {\bf 54}, 235 (1982).

\bibitem{Bax82}
R. J. Baxter, Exactly Solved Models in Statistical Mechanics (Academic Press,
  London, 1982), p. 363.

\bibitem{Car86}
J. L. Cardy, J. Phys. A: Math. Gen. 19, L1093 (1986).

\bibitem{Fuk89}
M. Fukugita and M. Okawa, Phys. Rev. Lett. {\bf 63}, 13 (1989).

\bibitem{Hu89}
C.-K. Hu and K.-S. Mak, Phys. Rev. B {\bf 40}, 5007 (1989).

\bibitem{Lee91}
J. Lee and J.M. Kosterlitz, Phys. Rev. B {\bf 43}, 1268 (1991).

\bibitem{Alv91}
N.A. Alves, B.A. Berg, and R. Villanova, Phys. Rev. B {\bf 43}, 5846 (1991).

\bibitem{Jan97}
Wolfhard Janke and Ramon Villanova, Nucl. Phys. B {\bf 489}, 679-696 (1997).

\bibitem{Sal97}
J. Salas and A. D. Sokal, J. Stat. Phys. 88, 567 (1997).

\bibitem{Dum17}
Hugo Duminil-Copin, Vladas Sidoravicius, Vincent Tassion Commun. Math. Phys.
  {\bf 349}, 47–107 (2017).

\bibitem{Jin12}
S. Jin, A. Sen, and A. W. Sandvik, Phys. Rev. Lett. {\bf 108}, 045702 (2012).

\bibitem{Jin13}
S. Jin, A. Sen, W. Guo, and A. W. Sandvik, Phys. Rev. B {\bf 87}, 144406
  (2013).

\bibitem{Pen24}
J.~H. Peng and F.-J. Jiang, Prog. Theor. Exp. Phys. {\bf 2024} 013A04.

\bibitem{BWTRG}
Adachi, Daiki and Okubo, Tsuyoshi and Todo, Synge, Phys. Rev. B {\bf 105},
  L060402 (2022).

\bibitem{TNS}
Zhao, H. H. and Xie, Z. Y. and Chen, Q. N. and Wei, Z. C. and Cai, J. W. and
  Xiang, T., Phys. Rev. B {\bf 81}, 174411 (2010).

\bibitem{Lev07}
M. Levin, C.P. Nave, Phys. Rev. Lett. {\bf 99}, 120601 (2007).

\bibitem{iTEBD}
Or\'us, R. and Vidal, G., Phys. Rev. B {\bf 78}, 155117 (2008).

\bibitem{CTMRG}
Or\'us, Rom\'an and Vidal, Guifr\'e, Phys. Rev. B {\bf 80}, 094403 (2009).

\bibitem{Xie12}
Z.Y. Xie, J. Chen, M.P. Qin, J.W. Zhu, L.P. Yang, T. Xiang, Phys. Rev. B {\bf
  86}, 045139 (2012).

\bibitem{ATRG}
Adachi, Daiki and Okubo, Tsuyoshi and Todo, Synge, Phys. Rev. B {\bf 102},
  054432 (2020).

\bibitem{CDL}
Gu, Zheng-Cheng and Wen, Xiao-Gang, Phys. Rev. B {\bf 80}, 155131 (2009).

\bibitem{Gilt}
Hauru, Markus and Delcamp, Clement and Mizera, Sebastian, Phys. Rev. B {\bf
  97}, 045111 (2018).

\bibitem{autodiff}
Liao, Hai-Jun and Liu, Jin-Guo and Wang, Lei and Xiang, Tao, Phys. Rev. X {\bf
  9}, 031041 (2019).

\bibitem{Bud93}
E. Buddenoir and S. Wallon, J. Phys. A: Math. Gen. {\bf 26}, 3045 (1993).

\bibitem{Ued23}
Atsushi Ueda and Masaki Oshikawa, Phys. Rev. B {\bf 108}, 024413 (2023).

\end{thebibliography}

\end{document}